\documentclass[12pt]{article}

\begin{document}

\begin{center}

{\Large {\bf I. Calculation of the observed value of large \\

\vspace{0,5cm}
mass hierarchy in modified RS model}}

\vspace{2cm}

{Boris L. Altshuler}\footnote[1]{E-mail adresses: altshuler@mtu-net.ru \& altshul@lpi.ru}

\vspace{0,5cm}

{\it Theoretical Physics Department, P.N. Lebedev Physical
Institute, \\  53 Leninsky Prospect, Moscow, 119991, Russia}

\vspace{2cm}

{\bf Abstract}
\end{center}

In generalized Randall-Sundrum (RS) model with dilaton where bulk potential is generated by the antisymmetric tensor field the mass term of this field
is introduced into the brane's Action. This
permits to stabilize brane's position and hence to calculate the
Planck/electroweek scales ratio which proves to depend
non-analytically on the dilaton-antisymmetric tensor field
coupling constant. The large observed number of mass hierarchy is
achieved for the moderate value of this coupling constant of order 0,3. In the subsequent Paper II it is shown that the same approach in a higher dimensional theory without dilaton permits to express mass hierarchy only through number of extra dimensions.
       
\newpage

\section{Introduction}

\quad To explain the existence in Nature on the fundamental level of the big number $M_{\rm Pl}/M=10^{16}$ ($M_{\rm Pl}$ - Planck mass, $M=1\rm TeV$ - electroweek scale) remains so far the challenge for theoreticians. The new insight to the problem was formulated in 1999 in five-dimensional Randall and Sundrum (RS) model~\cite{Randall} which essential ingredient - the possibility of low energy matter trapping on the brane - goes back to the 1983 Rubakov and Shaposhnikov pioneer paper~\cite{Rubakov}. In RS model because of the AdS-type exponential dependence of warp factor on the fifth coordinate, the large hierarchy number comes to much smaller number measuring the proper interbrane distance. The theory however remains non-predictive since the interbrane distance itself is not fixed by dynamics. The first attempt to use scalar field as a tool for brane stabilization was made by Goldberger and Wise~\cite{Goldb}, where however the branes' positions are fixed by the ad hoc form of scalar field potential. This situation is rather common: it is not a problem to introduce one or another dynamics capable to stabilize branes' positions, the problem is to do it in a natural way.

In this paper we do not introduce special "stabilizing" scalar
field and its potential but use most familiar scalar-gravity theory
containing $(p+2)$-form field strength,
$F_{p+2}=dA_{p+1}$, conformally coupled to the scalar field - dilaton;
strength of this interaction being measured by the dilaton coupling constant
$\alpha$. In $(p+2)$ dimensional space-time considered in this paper the nonzero field $F_{p+2}$ is conventionally a source
of the bulk cosmological term, here depending on dilaton.

The codimension one brane, being a boundary of space-time, must also "screen" the
potential $A_{p+1}$. The novel idea is to use for this screening
not Wess-Zumino term of a charged brane action which is linear in $A_{p+1}$, but mass term $\mu A_{p+1}^{2}$ "living" on a brane which is quadratic in $A_{p+1}$. This idea being accompanied with a natural
requirement of the same scaling behavior of the bulk and
brane terms of the action (which uniquely determines the dilaton coupling constants in the tension and mass
terms of brane action through the bulk dilaton coupling
$\alpha$) permits to express large hierarchy number
through small input dimensionless parameter $\alpha$ - see formula (\ref{21}) below (written for $p=3$). The non-analitical dependence on $\alpha$-squared in the RHS of (\ref{21}) is actually the main result of the paper.

The conventional Wess-Zumino term describing interaction of
charged $p$~-~brane with potential $A_{p+1}$ does not depend on
the induced brane's metric and hence does not contribute to the brane's
energy-momentum tensor. Contrary to this the "mass term" $\mu
A_{p+1}^{2}$ used in this paper contributes as a $\delta$-function source to both - to "Maxwell" equation for
potential $A_{p+1}$ and to Einstein's equations in space-time
directions parallel to the brane (resulting in the Israel jump conditions at the brane). In case the mass term $\mu A_{p+1}^{2}$ is the only term of the brane's action its pointed out "two-fold" role leads to discrepancy which looks strange - like $2=1$. The remedy is to introduce into the brane's action the
standard tension term; its strength $\sigma$ must be properly fine-tuned to
$\mu$.

Sec. 2 presents our primary action, dynamical equations and bulk solution of the model. In Sec. 3 Israel and other jump conditions at the brane are
written down and mechanism stabilizing brane's position is
demonstrated. In Sec. 4 mass hierarchy is calculated. In
conclusion possible trends of future work are outlined.

\section{Description of the model}

\quad We start from the Einstein frame action in $(p+2)$-dimensional space-time which is a "warped" product of flat Minkowski space-time $M^{1,p}$ and one extra dimension $z$. The action includes bulk terms and action of the p-brane limiting space-time in the extra direction $z$ "from above", i.e. $z<z_{\it{br}}$, with $Z_{2}$-symmetry imposed at $z=z_{\it{br}}$. (We shall note immediately that bulk solution presented below possesses singularity at $z=0$; this demands to limit space-time in $z$-direction also "from below" at some $z=z_{\it{min}}>0$; there are different ways to do it, the important point however is that the value of $z_{\it{min}}$ does not influence essentially the calculated value of mass hierarchy, see Sec. 4). Thus our action:

\begin{eqnarray}
\label{1}
S_{(p+2)}=M^{p}\int\Big\{R^{(p+2)}-\frac{\epsilon}{2(p+2)!}e^{-\alpha\phi}F_{p+2}^2-\frac{1}{2}(\nabla\phi)^2 \nonumber \\
-\frac{\hat\mu}{2(p+1)!} e^{-\beta\phi}A_{p+1}^{2}-\hat\sigma e^{\gamma\phi}\Big\}\sqrt{-g^{(p+2)}}\,d^{p+2}x+\rm{GH}.
\end{eqnarray}
where "Planck mass" $M$ in $(p+2)$ dimensions is
supposed to be of the electroweek scale; $R^{(p+2)}$ is a scalar curvature in $(p+2)$ dimensions, $g_{AB}$, $\phi$, $F$, $A$ denote metric, dilaton, $(p+2)$-form tensor field strength and its potential; $\rm{GH}$ - Gibbons-Hawking surface term; $\epsilon=\pm1$ (to receive generalization of the AdS-type behaviour typical for RS-model we are enforced to take below "abnormal" sign $\epsilon=-1$, whereas this is not the case in the higher dimensions model which will be considered in the Paper II). The dilaton coupling constants $\beta$, $\gamma$ in the tensor field mass term and tension term of the brane's action correspondingly are determined by the above mentioned demand of similar scaling behavior of all terms of the action (\ref{1}):

\begin{equation}
\label{2}
\beta=\alpha \, \frac{2p+1}{2p+2}, \qquad {}  \gamma=\alpha \, \frac{1}{2p+2}.
\end{equation}
These important relations also mean that constant shift of scalar
field $\phi$ has no impact upon solutions of dynamical equations
but just change multiplicatively $M$ in the action (\ref{1}); this permits to adjust $\mu=M$ in (\ref{1}), we'll use it later in Sec. 4.

$\hat\mu$ and $\hat\sigma$ in (\ref{1}) are densities located at the brane:

\begin{equation}
\label{3}
\hat\mu=\mu \frac{\delta(z-z_{\it{br}})}{N}, \qquad \hat\sigma=\sigma \frac{\delta(z-z_{\it{br}})}{N},
\end{equation}
where $N$ is a lapse function of $z$-coordinate transverse to the
brane. 

We take the standard anzats for the metric and antisymmetric $(p+1)$-form potential:

\begin{equation}
\label{4}
ds_{(p+2)}^{2}=b^{2}{\tilde g}_{\mu\nu}^{(p+1)}dx^{\mu}dx^{\nu}+N^{2}dz^{2},
\end{equation}
where $(p+1)$-dimensional metric ${\tilde g}_{\mu\nu}^{(p+1)}=\rm{diag}(-,+1,+1\cdots)$;

\begin{equation}
\label{5}
A_{p+1}=f(z)\epsilon_{\mu_{1}\cdots\mu_{p+1}},
\end{equation}
wherefrom

\begin{equation}
\label{6} F_{p+2}=f'(z)\epsilon_{\mu_{1}
\cdots\mu_{p+1}z}=Qe^{\alpha\phi}b^{p+1}N\epsilon_{\mu_{1}\cdots\mu_{p+1}z},
\end{equation}
prime means derivation over $z$, the last equality in (\ref{6}) is the solution of bulk "Maxwell" equation (\ref{12}), "charge" $Q$ is an arbitrary constant of the solution. From (\ref{5}), (\ref{6}) with account of (\ref{4}) and negative signature of metric ${\tilde g}_{\mu\nu}$ it follows:

\begin{equation}
\label{7}
\frac{A_{p+1}^2}{(p+1)!}=-\frac{f^2}{b^{2p+2}},
\end{equation}

\begin{equation}
\label{8}
\frac{F_{p+2}^2}{(p+2)!}=-\frac{f'^{2}}{b^{2p+2}N^{2}}=-e^{2\alpha\phi}Q^{2}.
\end{equation}

For this anzats the constraint and three second order dynamical equations for functions $b(z)$, $\phi(z)$, $f(z)$ follow in a standard way from the action (\ref{1}):

\begin{equation}
\label{9}
\frac{p(p+1)}{2N^2}\frac{b'^2}{b^2}=-\frac{\epsilon}{4}e^{-\alpha\phi}\frac{f'^{2}}{b^{2p+2}N^{2}}+\frac{\phi'^2}{4N^2};
\end{equation}

\begin{equation}
\label{10}
-\frac{1}{N^2}\left[\frac{b''}{b}-\frac{b'N'}{bN}+p\frac{b'^2}{b^2}\right]=\frac{\epsilon}{2p}e^{-\alpha\phi}\frac{f'^{2}}{b^{2p+2}N^{2}}+\frac{\hat\mu}{4p}e^{-\beta\phi}\frac{f^2}{b^{2p+2}}+\frac{\hat\sigma}{2p}e^{\gamma\phi};
\end{equation}

\begin{equation}
\label{11}
\frac{1}{N^2}\left[\phi''-\phi'\frac{N'}{N}+(p+1)\phi'\frac{b'}{b}\right]=\frac{\epsilon\,\alpha}{2}e^{-\alpha\phi}\frac{f'^{2}}{b^{2p+2}N^{2}}+\frac{\hat\mu\beta}{2}e^{-\beta\phi}\frac{f^{2}}{b^{2p+2}}+\hat\sigma\gamma e^{\gamma\phi};
\end{equation}

\begin{equation}
\label{12}
\frac{\epsilon}{b^{p+1}N}\left[\frac{b^{p+1}Ne^{-\alpha\phi}f'}{b^{2p+2}N^{2}}\right]'= \hat\mu e^{-\beta\phi} \frac{f}{b^{2p+2}}.
\end{equation}

We shall use the following bulk solution of these equations where we put $\epsilon=-1$ in (\ref{1}), (\ref{9})-(\ref{12}) and $z$ is taken to be a proper distance, i.e. $N=1$ in (\ref{4}):

\begin{equation}
\label{13}
b=\left(\frac{z}{l}\right)^{\xi}, \quad e^{\alpha\phi}=\left(\frac{z}{l}\right)^{-2}, \quad f=\frac{Ql}{[(p+1)\xi-1]}\left(\frac{z}{l}\right)^{[(p+1)\xi-1]}+ \rm const,
\end{equation}
where length $l$ and dimensionless constant $\xi$ are determined by $Q$ and $\alpha$:

\begin{equation}
\label{14}
\xi=\frac{2}{p\alpha^{2}}, \qquad Q^{2}l^{2}=\frac{4(2p+2-p\alpha^{2})}{p\alpha^{4}},
\end{equation}

The RS limit $\alpha \to 0$ of the solution (\ref{13}), (\ref{14}) is not immediately seen here (it is achieved if $l$ is taken $\sim \alpha^{-2} \to \infty$). However this limit becomes transparent if the bulk solution (\ref{13}), (\ref{14}) is rewritten e.g. in the lapse-gauge $N=b^{-1}$ in (\ref{4}); generalization of this form of the bulk solution is written down in Sec. 5.

Length $l$ which is an arbitrary dimensional constant of the solution, as well as dimensional constants in the action (\ref{1}), belong to "input" parameters of the theory. And following the general idea to have all "input" parameters of one and the same - electroweek - order we'll put $l=M^{-1}$ later on in formula (\ref{21}) determining Newton's constant. Also we shall discard the arbitrary $\rm constant$ in the expression for $f$ in (\ref{13}); because of the growth of $f$ with $z$ (for $\alpha^{2}<(2p+2)/p$, as it is seen from (\ref{13}), (\ref{14})) this $\rm constant$ will not influence essentially the calculated value of mass-scale hierarchy.

Now we shall consider the role of the "hatted" brane terms in the RHS of equations (\ref{10})-(\ref{12}).

\section{Jump conditions and stabilization of brane's position}

\quad Integrating Eqs. (\ref{10})-(\ref{12}) containing second derivatives $b''$, $\phi''$, $f''$ over $z$ around brane's position $z=z_{\it{br}}$, taking into account definitions (\ref{3}) and imposing $Z_{2}$ symmetry at the brane gives following Israel condition for $b'(z_{\it br})$ and analogous jump conditions for $\phi'(z_{\it br})$, $f'(z_{\it br})$:

\begin{equation}
\label{15}
\frac{2}{N^{2}} \frac{b'}{b}=\frac{\mu}{4pN} e^{-\beta\phi}\frac{f^{2}}{b^{2p+2}}+\frac{\sigma}{2pN} e^{\gamma\phi},
\end{equation}

\begin{equation}
\label{16}
-\frac{2}{N^{2}} \phi'=\frac{\mu\beta}{2N}e^{-\beta\phi}\frac{f^{2}}{b^{2p+2}}+\frac{\sigma\gamma}{N} e^{\gamma\phi},
\end{equation}

\begin{equation}
\label{17}
\frac{2}{N^{2}} e^{-\alpha\phi}f'=\frac{\mu}{N} e^{-\beta\phi}f.
\end{equation}
Eqs. (\ref{15})-(\ref{17}) are valid at $z=z_{\it br}$. Eq. (\ref{17}) is a crucial one wherefrom, after the substitution there of the solution (\ref{13}) (where we put $\rm const=0$ in the expression for $f$ - see discussion in the end of Sec. 2), the brane's position is determined through values of $\mu$, $l$, $p$, $\alpha$:

\begin{equation}
\label{18}
\frac{z_{\it br}}{l}=\left[\frac{2}{\mu l}\,\frac{(2p+2-p\alpha^{2})}{p\alpha^{2}}\right]^{(p+1)/p}.
\end{equation}

Substitution of this expression for $z_{\it{br}}$ and of the solution (\ref{13}) into other jump equations (\ref{15}), (\ref{16}) gives one and the same consistency condition:

\begin{equation}
\label{19}
2\mu=\mu+\frac{\sigma}{2p} (2p+2-p\alpha^{2}).
\end{equation}
As it was mentioned in the Introduction the "double" role of the mass term $\mu A_{p+1}^{2}$ in the equations of motion would curiously lead to $2=1$ discrepancy of the model in case we did not introduce brane's tension term. Eq. (\ref{19}) may be considered as a fine-tuning condition for brane's tension $\sigma$.

\section{Calculation of mass hierarchy in 4 dimensions}

\quad To calculate from the action (\ref{1}) the Planck/electroweek scale ratio (further on in this Section we shall put $(p+1)=4$) we must integrate over $z$ the "4-dimensional" term of the curvature $R^{(5)}$ in the action (\ref{1}) which is equal to $\tilde{R}^{(4)}/b^2$ (where $\tilde{R}^{(4)}$ is a scalar curvature in 4 dimensional space-time). Using metric (\ref{4}) specified in (\ref{13}) we get from (\ref{1}):

\begin{equation}
\label{20}
M_{\rm Pl}^{2}=M^{3}\int_{z_{\it{min}}}^{z_{\it{br}}}\left(\frac{z}{l}\right)^{2\xi}\,dz,
\end{equation}
where $\xi$ is given in (\ref{14}) and upper limit $z_{\it{br}}$ is given by Eq. (\ref{18}). As it is easily seen the choice of the lower limit of integral in (\ref{20}) does not effect the value of integral essentially; thus we put in (\ref{20}) $z_{\it {min}}=0$ and receive finally $M_{\rm Pl}/M$ as a function of dimensional quantities $M$, $\mu$, $l$ and dimensionless coupling constant $\alpha$ ($\alpha^{2}<8/3$):

\begin{equation}
\label{21}
\frac{M_{\rm{Pl}}}{M}=\left(\frac{2Ml}{3\delta}\right)^{1/2}\left[\frac{2}{\mu l}\left(\frac{8}{3\alpha^{2}}-1\right)\right]^{\delta}, \qquad \delta \equiv \frac{8}{9\alpha^{2}}+\frac{2}{3}.
\end{equation}

As it was already said above the "scaling" values of coupling constants (\ref{2}) permit to adjust $\mu=M$ in the action (\ref{1}); also it is possible to adjust $l=M^{-1}$ with the choice of arbitrary $(p+2)$-form charge $Q$ in the bulk solution (\ref{13}), (\ref{14}). Thus we take in (\ref{21}) the "electroweek" values of the dimensional constants $\mu$ and $l$:
\begin{equation}
\label{22}
\mu=l^{-1}=M=1\rm TeV.
\end{equation}
For this choice the observed value of $M_{\rm{Pl}}/M=10^{16}$ is calculated from (\ref{21}) for the dilaton coupling constant $\alpha \approx 0,3$.

\section{Discussion}

\quad In the approach considered above the calculation of large mass hierarchy comes after all to the search of physical grounds of a theory of type (\ref{1}) with moderate value of dilaton-tensor field coupling constant $\alpha$. In the next paper it will be shown however that in a theory without dilaton but with additional number of extra dimensions the same approach based upon introduction into the brane action of tensor field mass term permits to express the value of mass hierarchy through number of extra dimensions; in this case the observed value of mass hierarchy $10^{16}$ is achieved in particular in D13 space-time with $S^{7}$ and $S^{1}$ subspaces being added to the standard 5 dimensions of RS model.

Also it would be very interesting if some physical grounds for appearance in the brane action of the antisymmetric tensor field mass term were pointed out.

To conclude we'll present the natural generalization of space-time given by Eqs. (\ref{4}), (\ref{13}) which is also generalization of the known "brane-bolt" model~\cite{Louko},~\cite{Burgess} to the scalar-gravity theory. The problem of calculation of small positive cosmological term of the observable universe naturally arises in this context~\cite{Altshuler}. The model includes the additional compact space-like direction $y$ which plays role of Euclidian "time" in the solution presented below. Thus we consider the $(p+1)$-dimensional space-time in (\ref{4}) as a product of the Minkowski $p$-dimensional space-time and a circle $S^{1}$. The corresponding bulk solution in the scalar-gravity theory given by the action (\ref{1}) (where additional Maxwell field term was included in (\ref{1})) is described by the following generalization of the Reissner-Nordstrom metric:

\begin{equation}\label{23}
ds_{(p+2)}^{2}=b^{2}{\tilde g}_{\mu\nu}^{(p)}dx^{\mu}dx^{\nu}+\Delta dy^{2}+\frac{dr^2}{\Delta},
\end{equation}
where
\begin{equation}
\label{24}
b \sim r^{\kappa}, \quad \Delta (r)=C_{1}r^{2\kappa}+ C_{2}r^{1-p\kappa}+C_{3}r^{-2(p-1)\kappa}, \\
\qquad {} \kappa \equiv (1+p\alpha^{2}/2)^{-1}.
\end{equation}

Term $C_{1}$ in $\Delta$ is generated by the bulk potential $\sim F_{p+2}^{2}$; in case it is the only term in $\Delta$ Eqs. (\ref{23}), (\ref{24}) give the bulk solution (\ref{13}) rewritten in the lapse-gauge $N=b^{-1}$ (the RS-limit $\alpha=0$ is evident in this form of the solution). Term $C_{2}$ in $\Delta$ in (\ref{24}) generalizes the conventional Schwarzschild term, whereas term $C_{3}$ is a scalar-gravity generalization of the charged black-hole Raissner-Nordstrom term. The zero of $\Delta(r)$ at some $r=r_{0}$ called "bolt" regularly limits space-time which may be one of the tools to introduce lower limit in the integral (\ref{20}) determining Newton's constant.

It is evident that non-zero lower limit in the integral (\ref{20}) as well as non-zero $const$ in expression for $f$ in (\ref{13}) will give just a small corrections to the calculated value of Planck/electroweek scales ratio and to the brane's position (\ref{18}).
At the same time this will result also in the appearance of the additional relatively small terms in the consistency condition (\ref{19}); that means seemingly unnatural "fine-fine tuning" of brane's tension $\sigma$ in the primary action (\ref{1}).
However what is considered to be natural or unnatural essentially depends on one's taste and viewpoint. In the based on the AdS/CFT correspondence approach of "brain running" (see e.g. papers~\cite{Brevik} and references therein) brane action is not considered as a fundamental one but is obtained as a solution of dynamical equations (which in turn are equivalent to renormalization group flow equtions). Thus from this point of view fine-tuning of parameters of brane action demanded by dynamical equtions is not unnatural at all.

\section*{Acknowledgements} Athour is greatful to Vladimir Nechitailo for assistance. This work was partially supported by the grant LSS-1578.2003.2

\end{document}